\documentclass[11pt]{article}
\usepackage{latexsym,color,amsmath,amssymb,amsthm,fullpage}
\usepackage{graphicx}

\def \Exp{{\sf E}}

\def\CH{{\mathcal{CH}}}
\def\bridge{{\sf bridge}}
\def\U{{\mathcal U}}
\def\PT{{\cal PT}}
\def\CT{{\sf CT}}
\def\CV{{\sf CV}}
\def\CE{{\sf CE}}
\def\T{{\cal T}}

\def\prior{{\sf priority}}

\def\reals{{\mathbb R}}
\def\C{\mathcal{C}}

\def\U{\mathcal{U}}
\def\Q{\mathcal{Q}}
\def\L{{\cal L}}
\def\R{\mathcal{R}}

\def\prob{{\sf Pr}}

\def\Left{{\sf left}}
\def\Right{{\sf right}}
\def\apex{{\sf apex}}
\def\Mid{{\sf mid}}

\newtheorem{theorem}{Theorem}[section]
\newtheorem{lemma}[theorem]{Lemma}
\newtheorem{corollary}[theorem]{Corollary} 
\newtheorem{proposition}[theorem]{Proposition} 
\begin{document}

\begin{titlepage}
\title{A Kinetic Triangulation Scheme for Moving Points in The Plane\thanks{%
Work by Haim Kaplan and Natan Rubin was partially supported by Grant 975/06 from the
Israel Science Fund. Work by Micha Sharir and Natan Rubin was partially supported by
by Grants 155/05 and 338/09 from the Israel Science Fund. Work by Haim Kaplan was also supported by Grant 2006-204 from the U.S. - Israel Binational Science Foundation. Work by Micha Sharir was also supported by
NSF Grant CCF-08-30272, by grant 2006-194 from the U.S.-Israeli 
Binational Science Foundation, 
and by the Hermann Minkowski--MINERVA
Center for Geometry at Tel Aviv University.}}

\author{
Haim Kaplan\thanks{%
School of Computer Science, Tel Aviv University, Tel Aviv 69978,
Israel. 
E-mail: {\tt haimk@post.tau.ac.il }} \and
Natan Rubin\thanks{%
School of Computer Science, Tel Aviv University, Tel Aviv 69978,
Israel. 
E-mail: {\tt rubinnat@post.tau.ac.il }} \and
Micha Sharir\thanks{%
School of Computer Science, Tel Aviv University, Tel Aviv 69978,
Israel, and Courant Institute of Mathematical Sciences, New York
University, New York, NY 10012, USA. E-mail: {\tt
michas@post.tau.ac.il }} }
\maketitle

\begin{abstract}
We present a simple randomized scheme for triangulating a set $P$
of $n$ points in the plane, and construct a kinetic data structure 
which maintains the triangulation as the points of $P$ move
continuously along piecewise algebraic trajectories of constant
description complexity. Our triangulation scheme experiences an expected number of $O(n^2\beta_{s+2}(n)\log^2n)$ discrete changes, and handles them in a
manner that satisfies all the standard requirements from a kinetic
data structure: compactness, efficiency, locality and responsiveness. Here $s$ is the maximum number of times where any specific triple of points of $P$ can become collinear, $\beta_{s+2}(q)=\lambda_{s+2}(q)/q$, and $\lambda_{s+2}(q)$ is the maximum length of Davenport-Schinzel sequences of order $s+2$ on $n$ symbols. 
Thus, compared to the previous solution of Agarwal et al.~\cite{AWY}, we achieve a (slightly) improved bound on the number of discrete changes in the triangulation.
In addition, we believe that our scheme is simpler to implement and analyze.
\end{abstract}
\end{titlepage}

\section{Introduction}\label{Sec:intro}

Let $P(t)=\{p_1(t),\ldots,p_n(t)\}$ be a set of $n$ moving points 
in the plane. We assume that the motions of the points are simple, in
the sense that the trajectory of each point is a piecewise-algebraic
curve of {\em constant description complexity}, meaning that it can be
described as a Boolean combination of a constant number of polynomial
equalities and inequalities of constant maximum degree.

Our goal is to devise a reasonably simple scheme for triangulating 
$P(t)$ at any fixed time $t$, and to maintain the triangulation 
as the points move. That is, we wish to partition the 
convex hull $\CH(P)$ of $P$ into pairwise openly disjoint 
triangles whose vertices are the points of $P$, so that the 
interior of each triangle is empty---it does not contain any 
point of $P$. The scheme has to be {\em kinetic}, so that we 
can keep track of the discrete combinatorial changes that the 
triangulation undergoes as the points move, and update the
triangulation so that it continues to conform to the underlying
scheme. (That is, at any given time $t$ the maintained triangulation coincides with the one that would result in applying the static scheme to $P(t)$.)

The study of triangulations plays a central role in computational geometry because of
their numerous applications in such areas as computer graphics,
physical simulation, collision detection, and geographic information systems \cite{BernEpp, EdelsMeshes}. With the advancement in technology, many applications, for instance, video
games, virtual reality, dynamic simulations, and robotics, call for maintaining a triangulation as the
points move. 
 For example, the arbitrary Eulerian-Lagrangian method \cite{Lagrangean} provides a way to integrate
the motion of 
fluids and solids within a moving finite-element mesh.

In $\reals^2$, the Delaunay triangulation $DT(P)$ of $P$ produces well-shaped triangles, and thus 
is a good candidate for such a triangulation scheme.
The problem,
though, is that the best known upper bound on the number of 
discrete changes in $DT(P(t))$, as a function of time $t$,
is only nearly cubic in $n$ (the bound is cubic if the points move with constant velocities); see \cite{Stable,FU91,GMR91,SA:ds}. 
While it is strongly believed that the maximum possible number of 
discrete changes that $DT(P(t))$ can experience is only nearly 
quadratic in $n$, this is one of the hardest open problems in 
computational and combinatorial geometry (as recognized,
e.g., in \cite{TOPP}). 
Until this conjecture is established, one seeks alternative 
triangulation schemes with a provable \textit{nearly-quadratic} upper 
bound on the number of discrete changes. 
(This is best possible, since the convex hull itself can change $\Omega(n^2)$ times during a simple motion of the points of $P$; see \cite{SA:ds}.)
Moreover, the scheme 
should be sufficiently simple to define, to implement, and 
(as a secondary aesthetic virtue) to analyze. 
Finally, the 
scheme should satisfy the four basic properties of kinetic data 
structures \cite{BG99} detailed below.

Agarwal et al.~\cite{AWY} have recently presented such a randomized triangulation scheme which experiences
$O(n^22^{\sqrt{\log n \log\log n}})$ discrete changes. Their scheme, however, is fairly 
complicated, and its analysis is also rather involved. It uses a hierarchy of subsets $\emptyset=R_0\subseteq R_1\subseteq\cdots \subseteq R_w=P$, where each set $R_{i-1}$, for $1\leq i\leq w$, is a random sample of roughly $|R_{i}|^{1-1/i}\log n$ points of $R_{i}$. 
The algorithm maintains an entire hierarchy of triangulations $\emptyset=\T_0\subseteq \T_1\subseteq\cdots\subseteq\T_w=\T$, where each $\T_i$ is a triangulation of $R_i$; it is a refinement of $\T_{i-1}$ which is obtained by a suitable variant of the \textit{fan triangulation}, introduced in \cite{ABG}.

\medskip
\noindent{\bf Kinetic data structures.} 
The {\em Kinetic data structure\/} (KDS) framework, introduced by
Basch et al.~\cite{BG99}, proposes an algorithmic
approach, together with several quality criteria, for maintaining
certain geometric configurations determined by a set of objects,
each moving along a trajectory whose graph, as a function of time,
is a piecewise-algebraic curve (in space-time) of constant description complexity.
Several interesting algorithms have been designed, using this
framework, over the past decade, including algorithms for
maintaining the convex hull of a set of (moving) points in the
plane \cite{BG99}, the closest pair and all nearest neighbors
in any dimension \cite{AgKS,BG99}, and many other configurations. 
See \cite{Gui} for a comprehensive, albeit old, survey, and \cite{AgKS}
for a list of more recent results and references.

Typically, a KDS operates by maintaining a set of \textit{certificates}. As long as they are all valid, the structure being maintained is guaranteed to be valid too. Each certificate has a (first future) failure time, and we store these critical times in an event priority queue. When a certificate fails, we repair the KDS, update, if needed, the geometric structure that we maintain, generate new certificates and insert their failure times into the queue. 

Generally, a good KDS is expected to possess the following four properties: (i) \textit{Compactness}, meaning that the storage that it requires is larger only by a polylogarithmic factor than the space required for the structure being maintained. (ii) \textit{Efficiency}, meaning that the number of events that it processes (i.e., failure times of the certificates) is larger only by a polylogarithmic factor than the maximum possible number of discrete changes in the structure being maintained. (iii) \textit{Responsiveness}, meaning that repairing the KDS at a certificate failure event takes only  polylogarithmic time. (iv) \textit{Locality}, meaning that each input object is stored at only a polylogarithmic number of places in the KDS, so that an expected change in the motion of a single object can be processed efficiently. See \cite{AKS,BG99} for more details.

Therefore, a good KDS for 
kinetic triangulation in $\reals^2$ should have only nearly linear storage, process only a nearly-quadratic number of events, each in polylogarithmic time, and each moving point should be stored at only a polylogarithmic number of places in the KDS.

\medskip
\noindent{\bf Our result.}
In Section \ref{Sect:Static}, we present a simple triangulation scheme for a set $P$ of $n$ moving
points in the plane. For the sake of efficient kinetization we make
the scheme randomized, and assume a (natural) model in which the flight plans of
the moving points are independent of the randomization used by the
algorithm. The basic idea of the (static) triangulation is quite
simple (some details are glossed over in this informal overview):
We sort the points of $P$ by their $x$-coordinates, split $P$ at a
(random) point $p$ into a left portion $P_L$ and a right portion
$P_R$, compute recursively the upper convex hulls of
$P_L\cup\{p\}$ and of $P_R\cup\{p\}$, and merge them into 
the upper convex hull of the whole set $P$.

This process results in a {\em pseudo-triangulation} of the portion 
of the convex hull of $P$ lying above the $x$-monotone polygonal chain $\C(P)$ 
connecting the points of $P$ in their $x$-order. Each pseudo-triangle is
$x$-monotone, and consists of an upper {\em base} and of a left and
right lower concave chains, meeting at its bottom {\em apex}. See
Figure \ref{Fig:PseudoTail} for an illustration.
A symmetric process is applied to the portion of the hull below 
$\C(P)$, by computing recursively lower convex hulls of the
respective subsets of $P$.
(In particular, we obtain a hierarchical representation of $\CH(P)$, similar to the one of Overmars and van
Leeuwen~\cite{Overmars}; see also \cite{AKS}. See~\cite{ABG,TwoPolyg04,Speckmann} for additional applications of hierarchical pseudo-triangulations to kinetic problems.)

To obtain a proper triangulation of (the convex hull of) $P$, we partition each
pseudo-triangle $\tau$ into triangles. 
We accomplish this in the
following randomized incremental manner. We process the vertices
of $\tau$ (other than its apex and its leftmost and rightmost 
vertices) in order, according to the random ranks that they received
during the first splitting phase, and draw from each processed
vertex $v$ a chord, within the current sub-pseudo-triangle $\tau'$
of $\tau$ containing $v$, which splits $\tau'$ into two
sub-pseudo-triangles. This process ends with a triangulation of 
$\tau$, and we apply it to each of the pseudo-triangles, to obtain the full triangulation of $\CH(P)$.

In Section \ref{Sec:CombiChanges}, we prove that the expected number of events that
can arise during the motion is $O(n^2\beta_{s+2}(n)\log n)$ (with $s$ and $\beta$ as defined in the abstract), and that the expected number of discrete (also called topological) changes caused in our triangulation by each such event is bounded by $O(\log n)$.

In Section \ref{Sec:DataStruct}, we show how to maintain this triangulation, as the points 
of $P$ move, using a kinetic data structure that satisfies the criteria of \cite{BG99}, as listed above. 
There are several kinds of critical events we need to
watch for, in which pairs of points are swapped in the $x$-order or
triples of points become collinear.
We process each event of the former type in $O(\log^2n)$ expected time, and each event of the latter type in $O(\log n)$ expected time, for a total of $O(n^2\beta_{s+2}(n)\log^2n)$ (expected) processing time.
Our implementation encodes the pseudo-triangulation as a \textit{treap} on $P$ \cite{SA96}. 

The upper bounds that we obtain on the number of discrete events, and on their overall processing time, are slightly better than those of the scheme in \cite{AWY}, and we believe that our scheme is simpler (and more ``explicit") than that of \cite{AWY}.
\section{The Static Triangulation}\label{Sect:Static}
In this section we describe a simple scheme for constructing a static triangulation $\T(P)$ of $\CH(P)$.
We fix a random permutation $\pi$ of the points of $P$.  For each
$p\in P$ we denote its rank in $\pi$ as $\prior(p)$.  Let $\C(P)$
denote, as above, the $x$-monotone polygonal chain which connects the points of
$P$ in their $x$-order, assuming that no two points of $P$ have the
same $x$-coordinate. (In degenerate cases, which will arise at
discrete instances during the motion of the points of $P$, $\C(P)$
connects the points in the lexicographical order of their
coordinates.)  Since the two points of $P$ with extreme
$x$-coordinates are vertices of $\CH(P)$, $\C(P)$ partitions $\CH(P)$
into two components, $\CH^+(P)$ and $\CH^-(P)$, lying respectively
above and below $\C(P)$. With no loss of generality, we only describe
a triangulation $\T^+(P)$ of $\CH^+(P)$, and obtain the triangulation
$\T^-(P)$ of $\CH^-(P)$ in a fully symmetric fashion.  The overall
triangulation $\T(P)$ is the union of $\T^+(P)$ and $\T^-(P)$.

\medskip
\noindent
{\bf A static pseudo-triangulation of $\CH^+(P)$.}
We first construct a \textit{pseudo-triangulation} of $\CH^+(P)$ and
then refine it into a triangulation by partitioning each
pseudo-triangle into triangles.

\begin{figure}[htb]
\begin{center}
\input{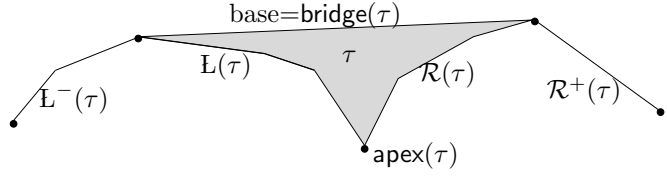}
\caption{\small\sf A single pseudo-triangle $\tau$ in our pseudo-triangulation of $\CH^+(P)$. In addition to its funnel (drawn shaded), $\tau$ has two tails $\L^-(\tau),\R^+(\tau)$.
 \label{Fig:PseudoTail}}
\end{center}
\end{figure}

\begin{figure}[htb]
\begin{center}
\input{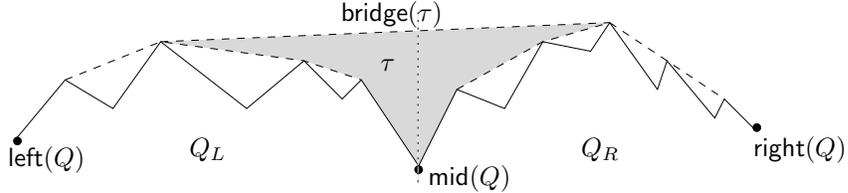}
\caption{\small\sf The recursive pseudo-triangulation of $\CH^+(Q)$. We add to $\PT^+(Q)$ the pseudo-triangle $\tau$ (whose funnel is drawn shaded), with endpoints
$\Left(\tau)=\Left(Q),\Right(\tau)=\Right(Q)$, and $\apex(\tau)=\Mid(Q)$, and then recursively construct $\PT^+(Q_L),\PT^+(Q_R)$. \label{Fig:PseudoTrian}}
\end{center}
\end{figure}

Each pseudo-triangle $\tau$ that we construct consists of a \textit{left tail}, a middle \textit{funnel}, and a \textit{right tail}
(any of these substructures may be empty; the tails were not mentioned in the overview in the introduction). The funnel
is an $x$-monotone simple polygon, whose
boundary consists of an upper \textit{base}, which is the segment connecting
its leftmost and rightmost vertices, and of a left and right lower concave
\textit{chains}, which are denoted respectively as $\L(\tau)$ and
$\R(\tau)$. The point in which $\L(\tau)$ and $\R(\tau)$  meet
is called  the \textit{apex} of $\tau$ and denoted
by $\apex(\tau)$. The left chain $\L(\tau)$ extends from the left endpoint of the base to $\apex(\tau)$, and the right chain extends from
$\apex(\tau)$ to the right endpoint of the base; see Figure
\ref{Fig:PseudoTail}.  
In addition, $\tau$ may have a left tail\footnote{\small These tailed pseudo-triangles are a special case of so-called \textit{geodesic triangles} introduced in \cite{Geodesic}.}
$\L^-(\tau)$ and a right tail $\R^+(\tau)$, so that $\L^-(\tau)$ is an $x$-monotone polygonal chain which
extends from the left vertex of the funnel to the left, till the {\em left
endpoint\/} $\Left(\tau)$ of $\tau$, so that $\L^-(\tau)\cup \L(\tau)$ is
a concave chain, and symmmetrically for $\R^+(\tau)$, which extends to
the right till the {\em right endpoint\/} $\Right(\tau)$ of $\tau$. Moreover, the line containing the base of $\tau$ is an upper common tangent of $\L^-(\tau)\cup \L(\tau)$ and $\R(\tau)\cup \R^+(\tau)$. Again, see
Figure \ref{Fig:PseudoTail}.
We construct the pseudo-triangulation of $\CH^+(P)$
recursively. At each step of the recursion we have some subset
$Q\subseteq P$ of points which are consecutive in the $x$-order
of $P$, and we construct a pseudo-triangulation $\PT^+(Q)$ of
$\CH^+(Q)$. At the topmost level of the recursion we have $Q=P$.
The construction of $\PT^+(Q)$ proceeds as follows; see Figure
\ref{Fig:PseudoTrian}.  Let $\Left(Q)$ (resp., $\Right(Q)$) denote the
point of $Q$ with the minimal (resp., maximal) $x$-coordinate, and let
$\Mid(Q)$ be the point $p$ of $Q\setminus\{\Left(Q),\Right(Q)\}$ with
the minimum value of $\prior(p)$.  Set $Q_L=\{p\in Q\mid x(p)\leq
x(\Mid(Q))\}$, $Q_R=\{p\in Q\mid x(p)\geq x(\Mid(Q))\}$ (so $\Mid(Q)$
belongs to both sets).  We add to $\PT^+(Q)$ the following pseudo-triangle
$\tau$. The base of $\tau$ is the
portion of the upper common tangent to $\CH^+(Q_L)$ and $\CH^+(Q_R)$ between the
points of tangency.
We call this base the {\em bridge\/} of $\tau$ and denote it by
 $\bridge(\tau)$.
The left (resp., right) chain $\L(\tau)$
(resp., $\R(\tau)$) is the portion of the upper hull of $Q_L$ (resp.,
$Q_R$) below  $\bridge(\tau)$.
 We take $\L^-(\tau)$ to be the portion of
the upper hull of $Q_L$ to the left of $\L(\tau)$, and define
$\R^+(\tau)$  symmetrically as the portion of the upper hull of
$Q_R$ to the right of $\R(\tau)$.  The points $\Left(Q)$ and
$\Right(Q)$ become the respective \textit{endpoints}
$\Left(\tau)$, $\Right(\tau)$ of $\tau$.  We also have
$\apex(\tau)=\Mid(Q)$ which belongs, by definition, to both chains. (The funnel of $\tau$
may be empty, if $\Mid(Q)$ is a vertex of the upper hull of $Q$. In
this case we can think of the funnel of $\tau$ as the singleton
$\apex(\tau)=\Mid(Q)$, and $\tau$ consists of the two tails
$\L^-(\tau),\R^+(\tau)$, meeting at $\Mid(Q)$, and forming together a common concave chain. Similarly, a pseudo-triangle may have an empty left tail and/or empty right tail.)

We then recursively pseudo-triangulate
each of $\CH^+(Q_L)$, $\CH^+(Q_R)$. The recursion
terminates when $|Q|\leq 3$ (by construction, $|Q|\geq 2$). If $|Q|=3$ then we output a single pseudo-triangle $\tau$, which is either a triangle,
when the midpoint lies below the segment connecting the endpoints, or, in the opposite case,
consists of the two segments $\L^-(\tau)=\Left(\tau)\apex(\tau)$ and
$\R^+(\tau)=\apex(\tau)\Right(\tau)$. If $|Q|=2$, no pseudo-triangle is output. In this case $\CH^+(Q)$ is a single edge of the chain $\C(P)$. 

Consider a pseudo-triangle $\tau$ such that 
$\Left(\tau)$ is not the leftmost point of $P$ and 
$\Right(\tau)$ is not the rightmost point of $P$.
Then one can show that the triple 
$(\Left(\tau),\Right(\tau),\apex(\tau))$ have the smallest priorities
among all points whose $x$-coordinates are between $x(\Left(\tau))$
and $x(\Right(\tau))$, inclusive (see Lemma \ref{Lemma:ConditionPseudo} below). To make this true for all 
pseudo-triangles, we
 augment the initial point set $P$ with two dummy points
$p_{-\infty}=(-\infty,-\infty)$ and $p_{\infty}=(\infty,-\infty)$, and
assign to them priorities $-1$ and $0$.
The upper hull of the augmented point set is
obtained from the upper hull of $P$ by adding two vertical downward-directed rays at the leftmost and rightmost points of $P$. Hence, any triangulation of $\CH^+(P)$ is also a triangulation of $\CH^+(P\cup\{p_{-\infty},p_{\infty}\})$, and vice versa.
In the rest of the paper we denote by $P$ the augmented point
set. 

The following lemma gives an operational definition of
$\PT^+(P)$, which will be used in the sequel.

\begin{lemma}\label{Lemma:ConditionPseudo}
  Let $a$, $b$, and $c$ be three points in $P$, such that
  $x(a)<x(b)<x(c)$. Then $\PT^+(P)$ contains a pseudo-triangle
  $\tau$ having endpoints $\Left(\tau)=a$, $\Right(\tau)=c$, and
  $\apex(\tau)=b$, if and only if\\ (i)
  $\prior(b)>\max\{\prior(a),\prior(c)\}$, and\\ 
  (ii) all points $p\in
  P$, such that $x(a)<x(p)<x(c)$ satisfy $\prior(p)\geq \prior(b)$.
\end{lemma}

\begin{proof}
  To prove the ``only if'' part we proceed by induction on our
  recursive construction.  Recall that at each recursive step we
  process some subset $Q\subseteq P$ whose points are
  consecutive in the $x$-order of $P$, and add to
  $\PT^+(P)$ a pseudo-triangle $\tau$ with
  $\Left(\tau)=\Left(Q)$, $\Right(\tau)=\Right(Q)$, and
  $\apex(\tau)=\Mid(Q)$.  To establish both asserted conditions (i)
  and (ii) for $\tau$, it is sufficient to observe that each point
  $p$, such that $x(\Left(Q)) < x(p) < x(\Right(Q))$, satisfies
  $\prior(p) > \max\{\prior(\Left(Q)),\prior(\Right(Q))\}$.  Indeed,
  the desired property holds initially for $P$ by our choice of
  the artificial points $p_{-\infty}$ and $p_{\infty}$ and their
  priorities. Assuming that this holds when we process some subset
  $Q$, and using the fact that $\Mid(Q)$ is the point with smallest
  priority in the range $x(\Left(Q)) < x(p) < x(\Right(Q))$, the claim
  also holds for $Q_L$ and $Q_R$.

  For the ``if'' part, we observe that for every choice of $b\in P$
  there is exactly one choice of $a$ and $c$ in $P$ so that the triple
  $(a,b,c)$ satisfies (i) and (ii), and every point $b\in P$ is an
  apex of exactly one pseudo-triangle of $\PT^+(P)$ (and the
  apex of each pseudo-triangle is distinct from each of
  $p_\infty$ and $p_{-\infty}$). 
  The latter is easy to establish by induction on the increasing order of the priorities of the points.
  This, combined with the arguments in the
  ``only if'' part, completes the proof.
\end{proof}

\medskip
\noindent
{\bf The pseudo-triangulation tree.}
The pseudo-triangulation $\PT^+(P)$ can be represented by a binary
tree in which every node $v$ represents a pseudo-triangle $\tau_v\in
\PT^+(P)$, and stores the point $p_v=\apex(\tau_v)$. The inorder
of the tree is the increasing $x$-order of the apices (i.e., the points of $P$). The subtree
rooted at $v$ represents the recursive pseudo-triangulation of
$\CH^+(P_v\cup\{\Left(\tau_v),\Right(\tau_v)\})$, where $P_v\subseteq
P$ denotes the set of points stored at the nodes of the subtree rooted
at $v$. Note that $\Left(\tau_v)$ and $\Right(\tau_v)$ are not stored
at this subtree---they are the next points to the left and to the
right of the points of $P_v$.
Abusing the notation slightly, we denote by $\PT^+(P)$ both the
pseudo-triangulation $\PT^+(P)$ and the tree
representing it.

\medskip
\noindent{\bf Remark:}
Let $v$ be a node in $\PT^+(P)$, so that $\Left(\tau_v)\neq p_{-\infty}$. Then $\Left(\tau_v)$ is stored at the lowest ancestor of $v$ whose right subree contains $v$. If $\Left(\tau_v)=p_{-\infty}$ then $v$ belongs to the path from the root of $\PT^+(P)$ to the leftmost leaf. Symmetric properties hold for $\Right(\tau_v)$.

In summary, we have the following lemma, whose proof is immediate from the construction.
\begin{lemma}\label{Lemma:treap}
  The tree representing $\PT^+(P)$ is a treap on $P\setminus\{p_{-\infty},p_{\infty}\}$. That is,
  $\PT^+(P)$ is a heap with respect to the \textit{priorities}, and
  a search tree with respect to the $x$-coordinates of the points.
\end{lemma}

\begin{figure}[htb]
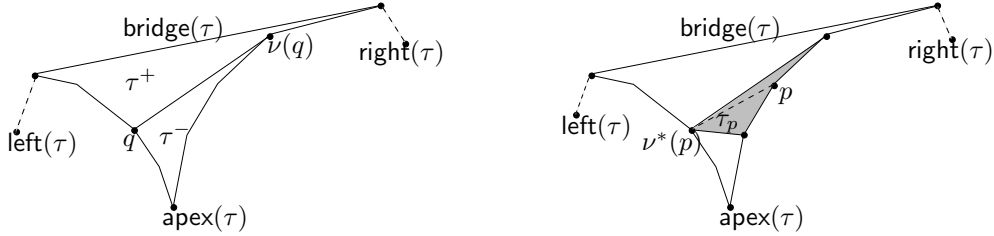

\begin{center}
\input{TriangulatePseudo2.pstex_t}\hspace{2cm}\input{TriangulatePseudo1.pstex_t}
\caption{\small\sf Left: The first step of triangulating a single pseudo-triangle $\tau\in \PT^+(P)$. Right: During the recursive construction of $\T(\tau)$ every non-corner vertex $p$ of the funnel of $\tau$ generates exactly one edge $e_p=p\nu^*(p)$, thus recursively splitting some sub-pseudo-triangle $\tau_p$ (drawn shaded). Note that in this figure $\nu^*(p)\neq \nu(p)$, which is the left endpoint of $\bridge(\tau)$.}\label{Fig:TrianPseudo}
\end{center}
\end{figure}

\noindent
{\bf Triangulating a fixed pseudo-triangle.}
Let $\tau$ be a pseudo-triangle of $\PT^+(P)$. Assume that the funnel of
$\tau$ is not empty, and is not already a triangle. We say that two
vertices $p,q$ of the funnel of $\tau$, where $p$ belongs to $\L(\tau)$ and $q$
belongs to $\R(\tau)$, are \textit{visible} from each other if $pq$
does not intersect $\partial \tau$ (except at its endpoints); in this
case $pq$ lies inside the funnel of $\tau$. Denote by $\nu(p)$ the
rightmost point on the right chain which is visible from
$p$. Note that either $\nu(p)$ is the
rightmost vertex of $\tau$ or $p\nu(p)$ is an upper tangent to
$\R(\tau)$. Symmetic definition and properties hold for points $q$ on
$\R(\tau)$. This definition also applies when $p$ is the leftmost vertex of
$\L(\tau)$ and when $q$ the rightmost vertex of $\R(\tau)$ (the endpoints of $\bridge(\tau)$), in which case $\nu(p)=q$ and $\nu(q)=p$. See Figure \ref{Fig:TrianPseudo} (left).


The triangulation $\T(\tau)$ of $\tau$ is obtained by recursively
splitting $\tau$ by chords into sub-pseudo-triangles, in the following
manner.  Choose the minimum priority vertex $q$ of the funnel of
$\tau$, other than the leftmost and the rightmost vertices and the
apex. Assume, without loss of generality, that $q$ lies on
$\L(\tau)$. See Figure \ref{Fig:TrianPseudo} (left). The segment $q\nu(q)$ splits $\tau$ into two
sub-pseudo-triangles $\tau^+$ and $\tau^-$. The pseudo-triangle $\tau^+$ has $q$ as an
apex and the same base as $\tau$. Its left chain is the portion of
$\L(\tau)$ from $q$ to the left, and its right chain is the
concatenation of $q\nu(q)$ with the portion of $\R(\tau)$ to the right
of $\nu(q)$. The pseudo-triangle $\tau^-$ has $q\nu(q)$ as its base, the
same apex as $\tau$, and its left and right chains are the portions of
$\L(\tau)$ and $\R(\tau)$ delimited respectively by $q$ and by
$\nu(q)$. A symmetric situation arises when $q\in \R(\tau)$. We add the
edge $q\nu(q)$ to $\T(\tau)$, and recursively triangulate each of
$\tau^+$ and $\tau^-$. We say that the edge $q\nu(q)$ in $\T(\tau)$ is
\textit{generated} by $q$.  In the further recursive steps, we redefine
$\nu(p)$, for vertices $p$ of each of these sub-pseudo-triangles,
restricting the visibility to only within the respective
pseudo-triangle. 
Note that for any pair of vertices $p,q$ that lie on the same chain of $\tau$, the segments $p\nu(p)$ and $q\nu(q)$ do not intersect in their relative interiors. Therefore, if $\nu(p)$ changes after a recursive call then it must change to a vertex of the base of the corresponding sub-pseudo-triangle.
See Figure \ref{Fig:TrianPseudo} (right). 
The recursion
bottoms out when the interior of $\tau$ is a triangle.
Note also that all the chords in $\T(\tau)$ cross the vertical ray above $\apex(\tau)$, and so they are totally ordered in the vertical direction.

\medskip
\noindent
{\bf Properties of $\T(\tau)$.}
Every vertex $p$ of the funnel of $\tau$, other than the leftmost and the rightmost vertices and the apex, generates exactly one edge $e_p$ during the whole recursive process. (For example, in Figure \ref{Fig:TrianPseudo} (left), the vertex $\nu(q)$ will not generate an edge in $\tau^-$, since it is an endpoint of that funnel, but will still generate an edge within $\tau^+$, or within some recursive sub-pseudo-triangle of $\tau^+$.) 
We denote by $\tau_p$ the sub-pseudo-triangle in which $e_p$ is generated, and by $\nu^*(p)$ the other endpoint of $e_p$. Note that $\nu^*(p)$ is either the original $\nu(p)$ or an endpoint of the base of $\tau_p$.





\section{Number of Discrete Changes in $\T(P)$}\label{Sec:CombiChanges}
In this section we bound the overall expected number of discrete changes that $\T(P(t))$ experiences as the points of $P$ move along (continuous) pseudo-algebraic trajectories of constant description complexity.
The analysis is with respect to a fixed random permutation $\pi$ of $P$ drawn ahead of the motion, so that the motion is ``oblivious'' to the choice of $\pi$. Thus, even though the $x$-order of the points may change during the motion, each point retains its initial priority, and the permutation $\pi$ is still a random permutation of $P$, with respect to the $x$-order of these points, at any fixed $t$.

\medskip
\noindent
{\bf Discrete changes in $\PT^+(P)$.}
For a fixed time instance $t\in \reals$, each pseudo-triangle $\tau\in
\PT^+(P(t))$ is defined by its endpoints 
$\Left(\tau)$, $\Right(\tau)$, and by its apex $\apex(\tau)$. 
Given such a triple of points, they define a
valid pseudo-triangle at time $t$ if and only if they, and the points
in-between in the $x$-order, satisfy the conditions of Lemma
\ref{Lemma:ConditionPseudo} (at time $t$).  Thus, as long as the
$x$-order of the points does not change, $\PT^+(P(t))$ does not change
either. That is, it consists of a fixed set of pseudo-triangles, each
defined by a fixed triple of points. However, the geometric structure of
a pseudo-triangle may change during such a time interval, and we will bound the number of these changes separately. Changes in (the labelings of the pseudo-triangles of) $\PT^+(P(t))$ occur only at discrete times when the
$x$-order of some pair of points in $P(t)$ changes; we refer to these
changes as $x$-\textit{swap events}.  

We assume that each
pseudo-triangle $\tau$ is present in $\PT^+(P(t))$ at a maximal connected time
interval $I(\tau)$, which is associated with $\tau$.  That is,
pseudo-triangles with the same triple $\Left(\tau)$, $\Right(\tau)$, and 
 $\apex(\tau)$  that appear in $\PT^+(P(t))$ at
disjoint time intervals, are considered distinct.  We emphasize that
all the other features of $\tau$, such as $\bridge(\tau)$, the
chains $\L(\tau)$ and $\R(\tau)$, and the triangulation $\T(\tau)$ of its funnel, may undergo discrete changes during
the time interval $I(\tau)$.  A pseudo-triangle $\tau$ is created or
destroyed only at a swap event when a point $p\in P$ with
$\prior(p)<\prior(\apex(\tau))$ crosses one of the vertical lines
through its endpoints $\Left(\tau)$ and $\Right(\tau)$ (of course, this also subsumes the cases where $\prior(p)$ is smaller than that of an endpoint of $\tau$), or when the
$x$-order of the points in the triple defining $\tau$ changes. In the
former case, if $\prior(p)>\max
\{\prior(\Left(\tau)),\prior(\Right(\tau))\}$ then $\tau$ is replaced
by another pseudo-triangle $\tau'$ with the same endpoints
$\Left(\tau')=\Left(\tau)$, $\Right(\tau')=\Right(\tau)$ but with $p$
as a new apex.

If $\prior(p)<\max\{\prior(\Left(\tau)),\prior(\Right(\tau))\}$ then $p$
replaces the endpoint it was swapped with.  Thus, each pseudo-triangle
$\tau$ in our kinetic pseudo-triangulation\\ $\PT^+(P(t))$ is
\textit{defined} by at most five points:
$\Left(\tau)$, $\Right(\tau)$, $\apex(\tau)$, and at most two additional
points which determine, by swaps with the endpoints of $\tau$, the
endpoints of the lifespan $I(\tau)$ of $\tau$ in $\PT^+(P(t))$.

\medskip
\noindent
{\bf Discrete changes in $\T(\tau)$.}
Fix a pseudo-triangle $\tau\in \PT^+(P(t))$. 
We only consider discrete changes in the funnel of $\tau$ and its triangulation $I(\tau)$, and ignore changes in the tails $\L^-(\tau), \R^+(\tau)$ (unless they also affect the funnel). This is because the changes in the tails will also show up as changes in the funnels of other pseudo-triangles that are created further down the recursion.

\begin{figure}[htb]
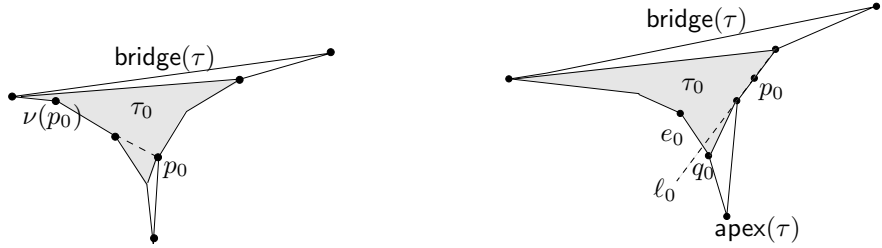

\begin{center}
\input{VisibilityEvent2.pstex_t} \hspace{2cm} \input{EnvelopeEvent.pstex_t}
\caption{\small\sf Left and right: visibility and envelope events (respectively). The sub-pseudo-triangle
$\tau_0$ contains all edges which are inserted to or deleted from $\T(\tau)$ at this event. \label{Fig:VisibilityEvent}}
\end{center}
\end{figure}

For a fixed time instance
$t\in I(\tau)$, the combinatorial structure of the triangulation
$\T(\tau)$ of $\tau$ depends only on the discrete structure of the
boundary of the funnel of $\tau$ (i.e., the ordered sequences of the
points along the chains $\L(\tau)$, $\R(\tau)$, and the base
$\bridge(\tau)$) and the visibility points $\nu(p)$ of all the
vertices of the funnel of $\tau$, excluding $\apex(\tau)$ (of course, it also depends on $\pi$).
Therefore, as the points of $P$ move during the time interval
$I(\tau)$, $\T(\tau)$ can change combinatorially only at events where the boundary
or visibility structure of $\tau$ changes. These events fall into the
following three types:

(i) \textit{Envelope events}, which occur at instances when one of the
chains $\L(\tau)$, $\R(\tau)$ contains three collinear vertices; see
Figure \ref{Fig:VisibilityEvent} (right).  This happens when a vertex
(which is not an endpoint of $\bridge(\tau)$) is added to or removed
from one of the chains bounding $\tau$.  We denote the total number of
such events during the period $I(\tau)$ by $E_\tau$.

(ii) \textit{Visibility events}, at which a vertex $q$ of $\R(\tau)$ becomes collinear with an edge $pr$ of $\L(\tau)\cup \L^-(\tau)$, or vice versa. See Figure \ref{Fig:VisibilityEvent} (left) ($\L^-(\tau)$ is relevant only for visibility events that affect $\bridge(\tau)$ and then $pr$ has to be its rightmost edge, and symmetrically for $\R^+(\tau)$).
This happens when $\nu(q)$ changes from $p$ to $r$, or vice versa. In particular, each (discrete) change of $\bridge(\tau)$ corresponds to a visibility event in which the bridge becomes collinear with an edge of $\L^-(\tau)\cup \L(\tau)$ or of $\R(\tau)\cup \R^+(\tau)$ that is incident to the respective endpoint of the bridge. We denote the total number of visibility events during $I(\tau)$ by $V_\tau$. 

A special case of this event occurs when $\bridge(\tau)$ is created (resp., destroyed), so that right before (resp., after) the event, the funnel of $\tau$ is empty. Note that immediately after (resp., before) the creation (resp., destruction) of $\bridge(\tau)$, the funnel of $\tau$ is a triangle.

(iii) \textit{Swap events}, at which some point $p\in P$, satisfying
$\prior(p)>\prior(\apex(\tau))$, crosses one of the vertical lines
through $\Left(\tau),\Right(\tau)$ or $\apex(\tau)$. Note that a
single swap event of this kind may cause massive discrete changes, of
highly unlocal nature, in the chains $\L(\tau)$, $\R(\tau)$, in the visibility
pointers $\nu(q)$ of the vertices of $\tau$, and in $\bridge(\tau)$.
See Figure \ref{Fig:EnvelopeEvent} (left) for an illustration.

Note that a swap between any other pair of points $p,q$ within the $x$-range of $\tau$ can be ignored in the present analysis, since the lower of the two points cannot belong to the funnel of $\tau$ at the time of swap.

Assuming general position of the trajectories of the points,
the above events occur at distinct time instances (except that the same event may show up, in different forms, in several pseudo-triangles).

A visibility event happens when $\nu(p)$ changes for some point $p$;
we then say that $p$ is {\em involved } in the visibility event.
An envelope event happens when a point $p$ joins or leaves one
of the chains $\L(\tau)$, $\R(\tau)$;
we then say that $p$ is {\em involved } in the envelope event.

\begin{lemma}\label{Lemma:ingoingEnvelopeEvent}
The only point $p$ for which $\nu(p)$ changes in an envelope event
is the point $p$ involved in the event.
\end{lemma}

\begin{proof}
 The lemma follows since at
  the moment following (resp., preceding) the appearance of $p$ on
  (resp., disappearance from) its chain, say $\L(\tau)$, its
  two incident edges are almost collinear. Thus, all vertices $q$ on
  the opposite chain satisfy $\nu(q)\neq p$ both before and
  after the event, and $\nu(q)$ is not affected by the event.
\end{proof}

\medskip
\noindent{\bf The number of changes in $\T(\tau)$.}
We define $P_\tau$ as the set of points $p\in P$, other than
$\apex(\tau)$, that appear on $\C(P)$ between $\Left(\tau)$ and
$\Right(\tau)$, at any time during the life span $I(\tau)$, and put
$N_\tau=|P_\tau|$. (Note that the points of $P_\tau$ may enter or leave the interval between $\Left(\tau)$ and $\Right(\tau)$ in the middle of $I(\tau)$, at $x$-swaps with either $\Left(\tau)$ or $\Right(\tau)$.) As noted above, every point $p\in P_\tau$ satisfies
$\prior(p)>\prior(\apex(\tau))$.  Clearly, our triangulation undergoes
$O(N_\tau)$ swap events during $I(\tau)$ (recall that we only consider
swaps with $\Left(\tau),\Right(\tau)$ or $\apex(\tau)$), and each of
them leads to $O(N_\tau)$ edge insertions and deletions to $\T(\tau)$
(the maximum number of edges in the whole triangulation $\T(\tau)$),
for a total of $O(N_\tau^2)$ such updates. We next bound the number of
discrete changes in $\T(\tau)$ caused by events of the remaining two
types.

Fix a set of at most five points that can potentially define a
pseudo-triangle for some set of priorities.  This set has an associated time interval $[t_1,t_2]$, and consists of
three points $a$, $b$, and $c$, such that, at all times $t_1<t<t_2$,
$x(a(t)) < x(b(t)) < x(c(t))$, and of two additional points $d_1$ and
$d_2$ (each of which could be equal to $b$), so that the $x$-coordinate of $d_i$ swaps
with either $a$ or $c$ at times $t_i$, for $i=1,2$. For some drawings of the random priorities, $\tau$ appears as 
a pseudo-triangle, and for other drawings it does not.
For $\tau$ to appear in $\PT^+(P(t))$, the priorities of 
$a=\Left(\tau)$ and
$c=\Right(\tau)$ should be smaller than the priority of 
$b=\apex(\tau)$. The priorities of $d_1$ and $d_2$ have to be at most the priority of $b=\apex(\tau)$,
and the priorities of all other points in  $P_\tau$ should be larger than
the priority of $b$. The probability of this to happen, assuming $a,b,c,d_1,d_2$ are all distinct, is easily seen to be $O(1/N_\tau^5)$ (for $N_\tau>0$).

When we
condition on drawings in which 
 $\tau$ indeed appears in  $\PT^+(P(t))$, the following holds.

\begin{proposition}\label{Thm:LogCostEvent}
  Let $\tau$ be a pseudo-triangle in the kinetic triangulation
  $\PT^+(P(t))$. Then the expected number of discrete
  changes in the triangulation $\T(\tau)$ of $\tau$, after any single
  envelope or visibility event which happens during the period $I(\tau)$, and conditioned on $\tau$ appearing in $\PT^+(P(t))$,
  is $O(\log N_\tau)=O(\log n)$.
\end{proposition}
\begin{proof}
  Clearly, the chords of $\T(\tau)$ (the additional edges which partition $\tau$ into triangles) admit a total vertical order, because
  they all cross the vertical line through $\apex(\tau)$.  Consider a
  time instance $t_0\in \reals$ when an envelope or a visibility event
  occurs, and let $p_0\in \L(\tau)\cup \R(\tau)$ be the point involved in
the event.
Let $t_0^-$ (resp., $t_0^+$) be the time right before (resp., after) the event. 
  Note
  that $p_0$ cannot be the apex of $\tau$ (unless the funnel of $\tau$ is already, or is going to become, a triangle). Note also that $p_0$ is not
  a vertex of $\bridge(\tau)$, neither at $t_0^+$ nor at $t_0^-$, unless
  $p_0$ is involved in a visibility event which changes
  $\bridge(\tau)$. In the latter case, $\T(\tau)$ gains or loses
 its topmost triangle at
  time $t_0$ but there are no other changes in the triangulation, as is easily checked.
 We may therefore assume that
  $\bridge(\tau)$ does not change at time
  $t_0$, and that $p_0$ is not a vertex of $\bridge(\tau)$.

  With no loss of generality, we assume that $p_0$ is a vertex of
  $\R(\tau)$ at time $t_0^+$, and treat the remaining cases
  symmetrically (for a visibility event, $p_0$ belongs to $\R(\tau)$ also at time $t_0^-$).  Consider the triangulation $\T(\tau)$ at time
  $t_0^+$ (that we would have obtained if we were to reconstruct $\T^+(P)$ statically at time $t_0^+$).  Let $\tau_0$ be the sub-pseudo-triangle of $\tau$ within which
  the edge $p_0\nu^*(p_0)$ is generated during the construction of $\T^+(P)$ (see Figure
  \ref{Fig:VisibilityEvent} (right)).  
  Note that the event at time $t_0$ leaves unchanged the visibility vertex $\nu(p)$ of each vertex $p$ in $\tau$ other than $p_0$.
  Indeed, this follows from Lemma \ref{Lemma:ingoingEnvelopeEvent} for envelope events and is obvious for visibility events, using our assumption that $\bridge(\tau)$ does not change. The recursive construction of $\T(\tau)$ is easily seen to imply that $\tau_0$
  appears as a sub-pseudo-triangle in the construction also at time $t_0^-$. Indeed, an easy inductive argument on the order of the ranks of the funnel vertices implies that the modified visibility vertices $\nu^*(p)$, and the resulting chords $p\nu^*(p)$, also do not change, up to the point where $\tau_0$ is constructed. Right after this step, the chord from $p_0$ is drawn, so the rest of the construction of $\T(\tau)$ might change completely, but only within $\tau_0$. Hence, $\tau_0$ contains every edge which
  is inserted to or deleted from $\T(\tau)$ at time $t_0$.  Therefore,
  the number of changes in $\T(\tau)$ is bounded by $O(W_0)$, where
  $W_0$ denotes the number of vertices of $\tau_0$ at the time of the event.

  Note that $W_0$ is a random variable depending (only) on the permutation
  $\pi(P_\tau)$ of $P_\tau$, which is obtained by restricting $\pi$ to
  $P_\tau$. Recall that we condition the analysis on permutations 
$\pi$ such that $\tau$ indeed appears in $\PT^+(P)$.
In these permutations, the points of $P_\tau$ have to follow all the (at most) five points defining $\tau$,
but as long as they obey this restriction
they can appear in any order.
It follows that, in our conditional probability subspace, the restriction
of $\pi$ to $P_\tau$ is a random permutation of  $P_\tau$.

To bound the expected value of $W_0$, we fix an arbitrary threshold $k\geq 10$ and prove that the event $\{W_0>k\}$ occurs with probability at most $O(1/k)$. The expected value of $W_0$ is then bounded
by 
\begin{equation}\label{Eq:LogCost}
\sum_{i=0}^{\log N_\tau} 2^{i+1}\prob\left\{W_0>2^i\right\}=O\left(\sum_{i=0}^{\log N_\tau} 1\right)=O(\log N_\tau).
\end{equation}

To show that $\prob\left\{W_0>k\right\}=O(1/k)$, we proceed through the following cases. In each case, except for the last one, we find a set $S_0$ of $\Omega(k)$ points which does not depend on $\pi(P_\tau)$, so that all its elements must appear in $\pi(P_\tau)$ after $p_0$. This readily implies the asserted bound. The last case is more involved but it is still based on the same general idea.

\medskip
\noindent {\it Visibility event.} 
If $\nu^*(p_0)$ is a vertex of the base of $\tau_0$, both at time $t_0^-$ and at time $t_0^+$, then $\T(\tau)$ does not change combinatorially at time $t_0$. Otherwise, as follows from the discussion in Section \ref{Sect:Static}, all three vertices that become collinear in the event appear in $\tau_0$, both before and after the event, which implies that $\nu(p_0)=\nu^*(p_0)$ at both times $t_0^-$ and $t_0^+$ (although they assume different values of these times).

Recall that $p$ is assumed to be a vertex of $\R(\tau)$, and suppose that $W_0>k$.  If $\tau_0$ contains at least $k/2$ vertices of
$\R(\tau)$, then it also contains a sequence $S_0$ of $k/4-1$
consecutive vertices of $\R(\tau)$ either immediately to the
left or immediately to the right of $p_0$.  Otherwise, $\tau_0$
contains $\nu(p_0)$ together with at least $k/2-1$ other vertices of
$\L(\tau)$, so it must contain a sequence $S_0$ of $k/4-1$
consecutive vertices of $\L(\tau)$ lying either immediately to the
left or immediately to the right of $\nu (p_0)$.  In both cases, the
key observation is that $S_0$ does not depend on
$\pi(P_\tau)$, and that $p_0$
precedes all the vertices of $S_0$ in $\pi(P_\tau)$ (except possibly for
one extremal vertex which is a corner of $\tau_0$). As noted above, this establishes the asserted bound.

\medskip
\noindent {\it Envelope event.} 
Again, suppose that $W_0>k$.  If $\tau_0$ contains
at least $k/2$ vertices of $\R(\tau)$, the bound follows by exactly the same argument as in the case of a
visibility event. 
Otherwise, if $\tau_0$ contains $\apex(\tau)$ we set $S_0$ to be the first $k/2-2$ points of $\L(\tau)$ to the left of $\apex(\tau)$.
Again, $S_0$ does not depend on $\pi(P_\tau)$, and all its elements must appear in $\pi(P_\tau)$ after $p_0$, so the bound follows.


We therefore assume that $\tau_0$ contains at most $k/2$ vertices of
$\R(\tau)$, and that its apex $q_0$ is distinct from $\apex(\tau)$.
Thus, the edge $q_0\nu^*(q_0)$ that $q_0$ generates is the lowest edge of $\tau_0$ which is a chord of $\tau$. We argue that
$\nu^*(q_0)=\nu(q_0)$ (before and after $t_0$; the definition of $\tau_0$ implies that $q_0$ precedes $p_0$ in $\pi(P_\tau)$). Indeed, otherwise, by the definition of
$\T(\tau)$, $\nu^*(q_0)$ is a vertex of the base of $\tau_0$, which
happens only if one of the chains of $\tau_0$ consists of the single
edge $q_0\nu^*(q_0)$. Since $p_0\in \R(\tau)$ and is involved in an envelope event, the edge $q_0\nu^*(q_0)$ must be the only edge of the left chain of $\tau_0$, which contradicts the fact that $\L(\tau)$ must contain at least $k/2$ vertices of
$\tau_0$ (for $k\geq 10$).  We distinguish between the following two cases.

(i) $q_0\in \L(\tau)$ (as depicted in Figure \ref{Fig:VisibilityEvent}
(right)). Then the entire left chain of $\tau_0$ is contained in $\L(\tau)$. Let $\ell_0$ be the line passing through $p_0$ and the
other two vertices of $\R(\tau)$ participating in the envelope event,
and let $e_0$ be the edge of $\L(\tau)$ intersected by
$\ell$. Clearly, $e_0$ is contained in $\tau_0$, because otherwise
$\R(\tau_0)$ would not be convex.
 If $\tau_0$
contains $k/4-1$ consecutive vertices of
$\L(\tau)$ which lie immediately to the left $e_0$, we set $S_0$ to be the set of these points, except for the leftmost one (which may be the endpoint of the base of $\tau_0$).
Otherwise we set $S_0$ to be the set of $k/4-2$ points lying on $\L(\tau)$ to the right of $e_0$. Since the definition of $e_0$ does not depend on $\pi(P_\tau)$, the set $S_0$ too does not depend on $\pi(P_\tau)$.

(ii) $q_0\in \R(\tau)$ (as depicted in Figure \ref{Fig:EnvelopeEvent} (right)). In this case we define at most $k/2$ sets, each consisting of $\Omega(1/k)$ points and independent of $\pi(P_\tau)$, such that all the points in at least one of these sets appear after both $p_0$ and $q_0$ in $\pi(P_\tau)$.
We fix $q_0$ on $\R(\tau)$ to the left of $p_0$ and define $S_{q_0}$
as the set of $k/2-2$ consecutive vertices of $\L(\tau)$ which appear at time $t_0$ (along $\L(\tau)$) immediately to the left of $\nu^*(q_0)=\nu(q_0)$. By the current assumptions, if $q_0$ is indeed the apex of $\tau_0$ then all points $q\in S_{q_0}$ belong to $\tau_0$ and, hence, satisfy $\prior(q)>\prior(p_0)>\prior(q_0)$. Since $q_0$ is fixed, $S_{q_0}$ is also fixed and is independent of $\pi(P_\tau)$. Hence, the above event happens with probability $O(1/k^2)$. 
Moreover, $q_0$ is one of the at most $k/2$ vertices of $\R(\tau)$ that lie to the left of $p_0$. Hence, by the probability union bound, the total probability of this scenario (over all the appropriate vertices $q_0\in \R(\tau)$) is $O(1/k)$.

\begin{figure}[htb]
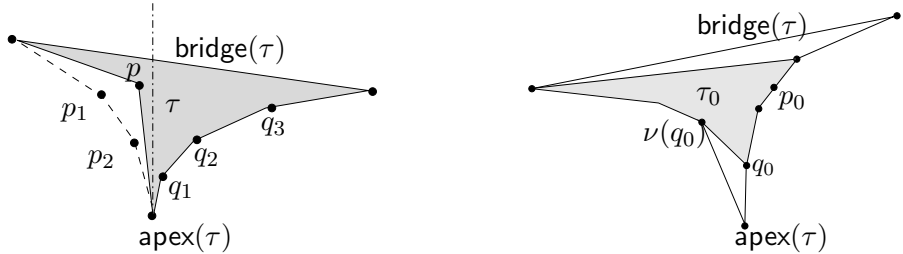

\begin{center}
\input{NonLocalSwap.pstex_t}\hspace{2cm}\input{EnvelopeEvent2.pstex_t}
\caption{\small\sf Left: Swap event. The funnel of $\tau$ immediately before the $x$-swap between $p$ and $\apex(\tau)$, which causes the vertices $p_1$ and $p_2$ to appear on $\L(\tau)$, and the vertices $q_1,q_2,q_3$ to disappear from $\R(\tau)$. Right: Envelope event. The case in which $q_0$ lies on $\R(\tau)$.} \label{Fig:EnvelopeEvent}
\end{center}
\end{figure}


We have proved that $\prob\left(W_0>k\right)=O(1/k)$ for any $k\geq 10$. This implies Equation (\ref{Eq:LogCost}) and completes the proof of Proposition \ref{Thm:LogCostEvent}.
\end{proof}

\begin{corollary}\label{Corol:LogCostEvent}
Let $\tau$ be a pseudo-triangle in the kinetic pseudo-triangulation $\PT^+(P(t))$. Then the expected number of edge insertions and deletions in $\T(\tau)$ during the period $I(\tau)$, conditioned upon the event that $\tau$ appears in $\PT^+(P)$, is $O((E_\tau+V_\tau)\log N_\tau+N_\tau^2)$.
\end{corollary}

For a fixed pseudo-triangle $\tau$ (including the choice of the connected life span $I(\tau)$), $V_\tau$ and $E_\tau$ are 2-valued random variables: They are $0$ if $\tau$ does not appear in $\PT^+(P)$, and assume a fixed ``deterministic" value if $\tau$ does appear.
The following theorem gives an upper bound on these values.


\begin{theorem}\label{Theorem:QuadEvents}
  For each pseudo-triangle $\tau $ we have
  $V_\tau=O(N_\tau^2\beta_{s+2}(N_\tau))$
and   $E_\tau=O(N_\tau^2\beta_{s+2}(N_\tau))$, where $s$ is the
  maximum number of times at which any fixed triple of points of $P$
  becomes collinear.
\end{theorem}
\begin{proof}
  We show the bound for visibility events. The bound for envelope
  events is known (see \cite{ABG,AKS}) and can be proved similarly.  

We fix a point $p\in
  P_\tau$ and count the number of visibility events where $p$ is a
  vertex of $\L(\tau)$ which is collinear with an edge of $\R(\tau)$.
  To do so, we define, for each $q\in P_\tau'=P_\tau\cup\{\Right(\tau)\}\setminus\{p\}$, a
  partially defined function $\varphi_{p,q}(t)$ which measures the
  angle between $pq$ and the $y$-axis, and whose domain consists of
  all $t\in \reals$ at which $x(\Left(\tau))\leq x(p)\leq
  x(\apex(\tau))\leq x(q)\leq x(\Right(\tau))$.  Clearly, each
  visibility event under consideration corresponds to a breakpoint of
  the lower envelope of $\{\varphi_{p,q}\}_{q\in P_\tau'}$ (but not necessarily vice versa; for example, such a breakpoint can arise at a time when $p$ is not a vertex of the funnel of $\tau$). Since any pair
  $\varphi_{p,q_1}$, $\varphi_{p,q_2}$ of these functions can intersect
  in at most $s$ points (these are times at which $p$, $q_1$, and $q_2$ are
  collinear),
and for each
$q$ the domain of  $\varphi_{p,q}(t)$ 
consists of a constant number of intervals (delimited by times at which either $p$ or $q$ swap with $\Left(\tau),\Right(\tau)$, or $\apex(\tau)$), it follows that the number of
  breakpoints is $O(N_\tau\beta_{s+2}(N_\tau))$ \cite{SA:ds}.  A
  symmetric argument holds for the number of visibility events where
  $p$ is a vertex of $\R(\tau)$ which is collinear with an edge of
  $\L(\tau)$. Repeating this analysis for each $p\in P$ yields the
  asserted overall bound.
\end{proof}

Fix a pseudo-triangle $\tau$. Conditioned on  priorities that
cause $\tau$ to appear in $\PT^+(P(t))$, Corollary \ref{Corol:LogCostEvent} and Theorem \ref{Theorem:QuadEvents} imply
that the expected number of discrete changes in $\T(\tau)$ is 
$O(N^2_\tau \beta_{s+2}(N_\tau) \log N_\tau)$. 
Let $\prob(\tau)$ be the probability that $\tau$ indeed
appears in $\PT^+(P(t))$. Then the total expected number of discrete 
changes in  $\PT^+(P(t))$
is

$$
 O\left(\sum_\tau \prob(\tau) N^2_\tau \beta_{s+2}(N_\tau) \log N_\tau\right)=
 O\left(\beta_{s+2}(n)\log n\sum_{\tau}\prob(\tau)N_\tau^2\right).
$$

\begin{lemma}\label{Lemma:BoundSumQuads} 
$\sum_\tau \prob(\tau) N^2_\tau = O(n^2\log n)$, where the sum is over all (possible sets of $1\leq h\leq 5$ points defining) possible pseudo-triangles $\tau$.
\end{lemma}

\begin{proof}
Without loss of generality, we only consider pseudo-triangles $\tau$ with $N_\tau>0$, which are defined by five distinct points of $P\setminus\{p_{-\infty}, p_{\infty}\}$. (Pseudo-triangles defined by fewer than five distinct points, or those
whose defining $5$-tuple includes $p_{-\infty}$ and/or $p_{\infty}$ are analyzed similarly, replacing the exponent $5$ by the appropriate $1\leq h\leq 4$.)
Thus, as already noted, $\prob(\tau)  = O(1/N_\tau^5)$, because $\tau$ appears in $\PT^+(P(t))$ if and only if the priorities of the
five points that define $\tau$ are smaller than the priorities of 
all other points in $P_\tau$
(and $\apex(\tau)$ has the largest priority among the defining points).
Therefore
$\displaystyle
\sum_\tau \prob(\tau)N_\tau^2 = O\left(\sum_\tau 1/N_\tau^3\right).$

In what follows, we call $N_\tau$ 
 the {\em level} of $\tau$.
Let $M_k(n)$ (resp.\ $M_{\le k}(n)$) denote the
maximum number of pseudo-triangles of level $k$ (resp., of level at most $k$), defined by $5$ points, in a set of
 $n$ moving points.
We claim that $M_{\leq k}(n)=O(n^2k^3)$. To see this, consider all the pseudo-triangles $\tau$ (defined by five points) whose birth time is determined by a fixed $x$-swap event occurring at some time $t_0$, between some pair of points $a,b\in P$. Assume without loss of generality that $a=\Left(\tau)$. Then $\apex(\tau)$ and $\Right(\tau)$ are among the $k+2$ points whose $x$-coordinates lie at time $t_0$ immediately to the right of $x(a)=x(b)$.
Similarly, the fifth point, which is responsible for the destruction of $\tau$, is one of the first $k+1$ points whose $x$-coordinates enter the interval between $x(\Left(\tau))=x(a)$ and $x(\Right(\tau))$. Thus, each of the $O(n^2)$ $x$-swap events defines the creation time of at most $O(k^3)$ pseudo-triangles of level at most $k$, which readily implies the asserted bound on $M_{\leq k}(n)$. We thus have
$$
\sum_\tau \prob(\tau) N^2_\tau = O\left(\sum_\tau 1/N_\tau^3\right)
= O\left(\sum_{k\geq 1} M_k(n)/k^3\right)
$$
$$
= O\left( \sum_{k\geq 1} M_{\le k}(n)/k^4 \right) =  O\left( \sum_{k\geq 1} n^2/k \right)  =  O(n^2\log n).
$$
\end{proof}


The combination of Corollary \ref{Corol:LogCostEvent},
Theorem 
\ref{Theorem:QuadEvents}, and Lemma \ref{Lemma:BoundSumQuads} implies
the following summary theorem.
\begin{theorem}
The total expected number of discrete changes in the kinetic triangulation $\T(P(t))$ is $O(n^2\beta_{s+2}(n)\log^2n)$.
\end{theorem}

\section{Kinetic Maintenance of $\T(P)$}\label{Sec:DataStruct}
In this section we describe a kinetic data structure which
supports efficient maintenance of $\T^+(P(t))$ under motion.
The structure satisfies\footnote{\small As in \cite{AKS}, all properties (except for compactness) hold \textit{in expectation}, with respect to the random permutation $\pi$.} the standard requirements of \textit{efficiency, compactness, responsiveness}, and \textit{locality}, as reviewed in the introduction.

\medskip
\noindent
{\bf The static structure.}
We store the pseudo-triangulation tree $\PT^+(P)$ as a treap over $P$, as
described in Theorem \ref{Lemma:treap}, whose inorder is the $x$-order of the points and where
the heap order is according to their random priorities. Each node $v$ in $\PT^+(P)$
corresponds to the pseudo-triangle $\tau_v$ 
whose apex is the point stored at $v$. We also store at $v$, as auxiliary data,
the endpoints $\Left(\tau_v)$ and $\Right(\tau_v)$, which are inherited from appropriate ancestors of $v$.

In addition, we also store at $v$ the combinatorial description of the
funnel of $\tau_v$, and of its triangulation $\T(\tau_v)$.  This includes
$\bridge(\tau_v)$, two ordered lists storing the vertices of
$\L(\tau_v)$, and $\R(\tau_v)$ in their left-to-right order, and the list of
the chords of $\T(\tau_v)$, sorted in their vertical order (i.e., the order of their intersections
with the vertical line through $\apex(\tau_v)$).  We
represent any sorted list of vertices or edges\footnote{\small Note that we do not store explicitly the tails $\L^-(\tau),\R^+(\tau)$, because the overall storage that they would require could be too large, as they can be shared by many pseudo-triangles.} as a balanced
binary tree supporting each of the operations {\sf search}, {\sf split}, and {\sf concatenate},
 in $O(\log n)$ time \cite{Tarjan}.
To facilitate efficient kinetic maintenance of $\T^+(P(t))$, we also store
the vertices of the upper hull of $P$, in their left-to-right
order in a balanced search tree. Note that each edge of the triangulation (not on $\C(P)$) appears twice in our structure, once as $\bridge(\tau_v)$ for some pseudo triangle $\tau_v$, and once on $\L(\tau_w)$ or $\R(\tau_w)$ for some ancestor $w$ of $v$ or on the convex hull of $P$.

\begin{theorem}\label{Thm:InitStructure}
  Let $P$ be a set of $n$ points in the plane.  The
  pseudo-triangulation tree $\PT^+(P)$, augmented with the auxiliary data items,
  as above, uses $O(n)$ space, and it can be initialized in $O(n\log
  n)$ time.
\end{theorem}
\begin{proof}
  The asserted bound on the overall storage follows from the easy
  observation that $\PT^+(P)$ contains $O(n)$ nodes, and every point
  $p\in P$ appears as a non-corner vertex on at most one chain
  $\L(\tau_v)$, $\R(\tau_v)$, over all nodes $v$ of $\PT^+(P)$.

  We construct the pseudo-triangulation tree $\PT^+(P)$ (excluding the
  auxiliary items $\bridge(\tau_v)$, $\L(\tau_v)$, $\R(\tau_v)$ and
  the chords of $\T(\tau_v)$) in a single \textit{top-down} pass, which implements the recursive construction given in Section \ref{Sect:Static}.  Clearly, this can be done
  in $O(n)$ time, after an initial sorting of the points of $P$, by their
  $x$-coordinates and by their priorities; sorting the points
  takes
  $O(n\log n)$ time.

  We next compute the items $\L(\tau_v)$, $\R(\tau_v)$, and $\bridge(\tau_v)$
  stored at the nodes $v$ of $\PT^+(P)$, by a single \textit{bottom-up}
  traversal of $\PT^+(P)$, which computes for every node $v$ the upper
  hull $\U(v)$ of the set $P_v\cup \{\Left(\tau_v),\Right(\tau_v)\}$.
  When we process a new non-leaf node $v$, we have already visited
  its respective left and right children $v_\ell$ and $v_r$, so their hulls $\U(v_\ell)$ and $\U(v_r)$ are already available.  We compute
  $\bridge(\tau_v)$ in $O(\log n)$ time by a simultaneous binary seach
  over $\U(v_\ell)$ and $\U(v_r)$, in the manner described in
  \cite{Overmars}. Then we use  $\bridge(\tau_v)$
  to split $\U(v_\ell)$ (resp.,
  $\U(v_r)$) into $\L^-(\tau_v)$ and $\L(\tau_v)$ (resp., $\R(\tau_v)$
  and $\R^+(\tau_v)$).  We store explicitly the chains $\L(\tau_v)$ $\R(\tau_v)$ at
  $v$, and compute $\U(v)$ by concatenating the three edge lists
  $\L^-(\tau)$, $\{\bridge(\tau)\}$, and $\R^+(\tau)$, in a similar manner to that described in \cite{AKS}. Overall, we spend
  $O(\log n)$ time at each node of $\PT^+(P)$, for a total of
  $O(n\log n)$ time.

  Finally, for each node $v$ in $\PT^+(P)$, we compute the list of chords of
  $\T(\tau_v)$ using the recursive mechanism described in Section \ref{Sect:Static}.
  Recall that every non-corner vertex $p$ of the funnel of $\tau_v$ generates
  exactly one edge $e_p$ which recursively splits the unique
  sub-pseudo-triangle $\tau_p$ of $\tau_v$. We process the non-corner
  vertices of $\T(\tau_v)$ in the increasing order of their
  priorities, and store the edges constructed so far in a list, in
  the order of their intersections with the vertical line through
  $\apex(\tau_v)$.

  It takes $O(\log n)$ time to process a non-corner vertex $p$ of
  $\tau_v$, for a total of $O(n\log n)$ time. Indeed, we can determine the
  corners of $\tau_p$ in $O(\log n)$ time, by a binary search over the
  list of the previously generated edges. In addition, we can
  determine $\nu(p)$ by a binary search over the appropriate chain
  $\L(\tau_v)$ or $\R(\tau_v)$, obtain $\nu^*(p)$ in $O(1)$ additional time, and insert the chord $p\nu^*(p)$ into the list of chords in $O(\log n)$ time.
\end{proof}

\medskip
\noindent
{\bf The kinetic certificates.}
To ensure the validity of $\PT^+(P)$ and its triangulation $\T^+(P)$,
we use three types of \textit{certificates}, denoted as $\CT$, $\CE$ and
$\CV$. Each certificate is a  predicate on  a constant number of
points.  As long as all the certificates remain true, the validity of
$\PT^+(P)$ and $\T^+(P)$ is ensured.  Each certificate contributes a
critical event to the global event priority queue $\Q$, which is the first
future time at which the certificate becomes invalid (if there is such a
time).

\smallskip
\noindent {\bf \emph{$\CT$-certificates}.} To ensure the validity of the tree
$\PT^+(P)$ (ignoring the auxiliary data), each pair of points $p,q\in
P$ with consecutive $x$-coordinates contributes a $\CT$-certificate
asserting that the order of $x(p)$ and $x(q)$ remains unchanged.
This certificate fails at the first future moment of an $x$-swap between $p$ and $q$.
According to Lemma \ref{Lemma:ConditionPseudo}, $\CT$-certificates (together with the chosen priorities) are
sufficient to ensure the validity of the ``bare" tree $\PT^+(P)$.

\smallskip
\noindent {\bf \emph{$\CE$-certificates}.} 
For each node $v$ in $\PT^+(P)$, the edge $\bridge(\tau_v)=pq$  contributes a
$\CE$-certificate ensuring that the (current) neighbors of $p$ and $q$
on $\L^-(\tau_v)\cup \L(\tau_v)$ and $\R(\tau_v)\cup \R^+(\tau_v)$
remain below the line through $p$ and $q$.  This certificate
involves\footnote{\small If $\bridge(\tau_v)$ does not exist then we have an
  even simpler certificate which fails when the two edges of
  $\L^-(\tau_v),\R^+(\tau_v)$ incident to $\apex(\tau_v)$ become
  collinear.} at most six points and fails at the first future time of
collinearity between $p,q$, and one of their four neighbor vertices on
$\L^-(\tau_v)\cup \L(\tau_v)$ and on $\R(\tau_v)\cup \R^+(\tau_v)$.

So far, we have ensured the validity of the tree $\PT^+(P)$ and of the edges $\bridge(\tau_v)$ stored at its nodes $v$. Moreover, the validity of all the chains $\L(\tau_v),\R(\tau_v)$ is also ensured because each one of their edges either belongs to $\C(P)$ or appears as $\bridge(\tau_w)$ at some descendant $w$ of $v$. Here a collinearity between three consecutive points on $\L(\tau_v)$ or on $\R(\tau_v)$ (an envelope event) will be detected as a change in $\bridge(\tau_w)$, for the appropriate descendant $w$. 
Similarly, the validity of the upper hull of $P$ follows since each of its edges either belongs to $\C(P)$ or appears as $\bridge(\tau_v)$ at some node $v$.
See \cite{AKS} and \cite{Overmars} for more details.

\smallskip
\noindent {\bf \emph{$\CV$-certificates}.}
It only remains to ensure the validity of the triangulations $\T(\tau_v)$, over all nodes
$v\in \PT^+(P)$. For this we need the third type of certificates, denoted by
$\CV$.  Fix a node $v$ in $\PT^+(P)$. Every internal point $p$ of
$\L(\tau_v)$ or $\R(\tau_v)$ contributes a $\CV$ certificate ensuring
the validity of $\nu(p)$. This certificate involves $p$, $\nu(p)$, and 
the two points adjacent to 
 $\nu(p)$ on its chain. It fails when one
of the points adjacent to
$\nu(p)$  becomes collinear with $p$ and $\nu(p)$.

Clearly, all of the above certificates use $O(n)$ storage,
and can be initialized, including the construction of the event queue $Q$ of their first failure times, by the algorithm of Theorem
\ref{Thm:InitStructure}, without increasing its overall assymptotic
running time, i.e., in $O(n\log n)$ time.

\medskip
\noindent
{\bf Handing critical events.} 
We next describe the repair operations 
required when an event, at which some certificate fails, happens. 

\smallskip
\noindent{\bf \emph{$\CT$-certificates.}} Failure of a $\CT$-certificate occurs at an $x$-swap.
That is,
the order of the $x$-coordinates of two consecutive points along $\C(P)$ switches, at some time $t=t_0$.

With no loss of generality we assume that $\prior(p)<\prior(q)$,
implying that $q(t_0^-)$ is a descendant of $p(t_0^-)$, where $t_0^-,t_0^+$ denote the time just before and just after $t_0$, respectively.  To update
 $\PT^+(P(t_0^+))$ we reconstruct  from scratch the subtree rooted at the node $v$ 
containing $p$, and recompute the kinetic
 certificates associated with its nodes and the points that they
contain. We remove the failure times
 of the expired certificates from $\Q$, and insert the new ones.  All
 this can be done in $O(n_v\log n_v)$ time using the algorithm of
 Theorem \ref{Thm:InitStructure}, where $n_v=|P_v|$. We prove that $\Exp \{n_v\}=O(\log n)$ by applying a
 simplified version of the analysis used in Proposition
 \ref{Thm:LogCostEvent}. As above, it suffices to show that
 $\prob\{n_v>k\}\leq 4/k$, for any $k\geq 1$.  Indeed, $n_v>k$ implies
 that either each of the $k/2$ points $w$
whose $x$-coordinates immediately precede $x(p)$ or each of the
$k/2$ points $w$
whose $x$-coordinates immediately follow $x(p)$ at time $t_0$
satisfies  $\prior(p)<\prior(w)$. 
This happens\footnote{\small We emphasize again that arguments of this kind are based on the assumption that the motion of the points is oblivious to the choice of priorities.} with probability at most
 $4/k$. Thus, we can reconstruct the subtree rooted at $v$ in $O(\log
 n\log\log n)$ expected time.

\begin{figure}[htb]
\begin{center}
\input{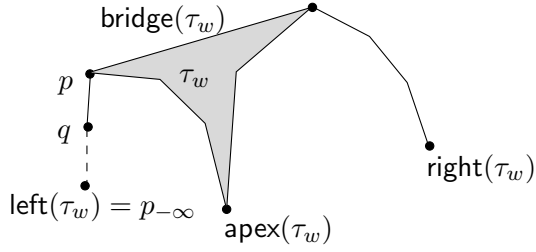}
\caption{\small\sf The view after a swap event between a pair of points $p,q$ with the smallest $x$-coordinates stored in the left subtree of a node $w$, whose $\CE$-certificate has to be updated.}
 \label{Fig:SwapEnvelope}
\end{center}
\end{figure}

 As can easily be checked, if neither $p$ nor $q$ is
the leftmost or the rightmost point of $P$ (excluding the points
at infinity which we added)
 then no further updates outside the
 subtree of $v$ are needed, and no additional certificates need to be
 created or destroyed. (That is because $\Left(\tau_v)\neq p_{-\infty}$ and $\Right(\tau_v)\neq p_{\infty}$, so the upper hull $\U(v)$ contains at most one of $p,q$, and it does not change as a result of the swap.)
 We next describe the necessary modifications in the setting, depicted in Figure \ref{Fig:SwapEnvelope}, in which case we
 assume that (i) $p$ and $q$ are the two points with the smallest
 $x$-coordinates, (ii) $x(q(t_0^-))>x(p(t_0^-))$, and (iii)
the $y$-coordinate of
 $p$ is larger than at $q$
when they swap; the other cases are treated
 symmetrically.  The $x$-swap between $p$ and $q$ causes $q$ to appear
 on the upper hull of $P$, below and to the left of $p$.
We add $q$ to the upper hull in  $O(\log n)$
 time. Similarly, $q$ becomes part of the tail $\L^-(\tau_w)$ of every
 ancestor $w$ of $v$ (both $w$ and $v$ lie on the leftmost path of the treap). If $w$ is such an ancestor whose bridge is incident to $p$ (from the right), then we have to incorporate $q$ into the certificate of
 $\bridge(\tau_w)$, and possibly replace its old failure time in
 $\Q$ with a new one. Since the expected number of ancestors $w$ of
 $v$, in the treap $\PT^+(P)$, is $O(\log n)$ (see, e.g., \cite{SA96}), any swap event can
 be processed in $O(\log^2n)$ expected time.

\begin{figure}[htb]
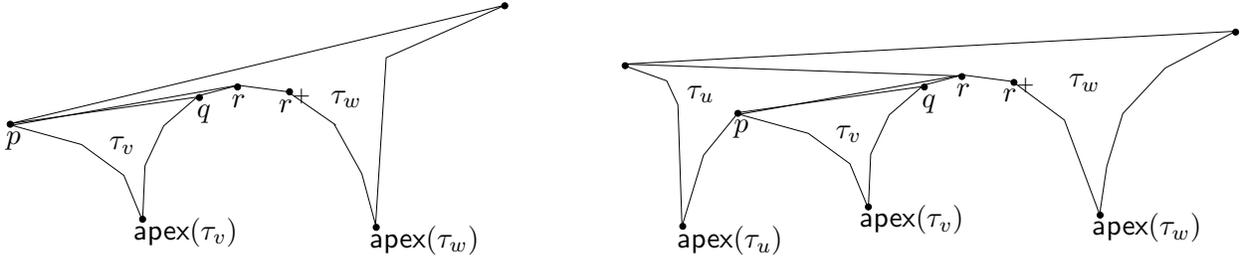

\begin{center}
\input{VisibilityEnvelope.pstex_t}\hspace{1.5cm}\input{VisibilityEnvelope1.pstex_t}
\caption{\small\sf Failure of the $\CE$-certificate at $v$ (shown at time $t_0^+$ right after the event). Left: The case where the ancestor $w$ that stores $rr^+$ coincides with the ancestor $u$ that has lost $q$. Right: The case where $u$ is distinct from $w$.}
 \label{Fig:UpdateAncestor}
\end{center}
\end{figure}

\smallskip
\noindent{\bf \emph{$\CE$-certificates}.} 
Consider a time $t_0$ when a 
$\CE$-certificate at some node $v$
fails.
We assume
without loss of generality that at time $t_0$ the leftmost vertex $p$ of $\L(\tau_v)$ becomes collinear with the leftmost edge $qr$ of $\R^+(\tau_v)$, so that $\bridge(\tau_v)$ was $pq$ before the event and is $pr$ afterwards, and
treat the remaining cases symmetrically. See Figure \ref{Fig:UpdateAncestor} for an illustration. As a result of this event,
the edge $pr$ replaces $pq$ as $\bridge(\tau_v)$, the edge $qr$ is
added to the end of $\R(\tau_v)$, and the triangulation $\T(\tau_v)$ gains the new
triangle $\triangle pqr$.  We need $O(\log n)$ time to update the
edge lists of $\R(\tau_v)$ and  $\T(\tau_v)$, and to compute the
$\CV$-certificate of $q$ (which ceases to be the endpoint of
$R(\tau_v)$) and add its failure time to $\Q$. (Note that the $\CV$-certificate of $q$ is part of the former $\CE$-certificate at $v$.)

To recompute the new certificate of $\bridge(\tau_v)$, we have
to determine the next edge $rr^+$ of $\R^+(\tau_v)$ that is incident to $r$
from the right.  This edge is either stored in one of the lists $\L(\tau_w)$
or $\R(\tau_w)$ at some ancestor $w$ of $v$, or it belongs to the
upper hull of $P$. See Figure \ref{Fig:UpdateAncestor} (left).
We find $rr^+$ by doing a binary search on the lists $\L(\tau_w)$ and $\R(\tau_w)$ for the ancestors $w$ of $v$, and if necessary also on the convex hull of $P$.

If $pq$ and $qr$ were part of the upper hull at time $t_0^-$, we replace them by a single edge $pr$, in $O(\log n)$ time.
Otherwise, $v$ has some ancestor $u$ such that $pq$ and $qr$ are stored in the edge list of
$\L(\tau_{u})$ or $\R(\tau_{u})$. (There is exactly one such ancestor $u$, which is equal to $w$ unless $r$ is incident to $\bridge(\tau_u)$; see Figure \ref{Fig:UpdateAncestor}. In the terminology of Section \ref{Sec:CombiChanges}, $\tau_u$ experiences an envelope event at time $t_0$.)
We find $u$ in $O(\log^2n)$ expected time by searching the edge lists stored at all ancestor nodes of $v$, whose expected number is bounded by $O(\log n)$. We then replace $pq$ and $qr$ by $pr$ in the edge list of the respective chain $\L(\tau_u)$ or $\R(\tau_u)$, and remove from $\Q$ the failure
time of the $\CV$-certificate of $q$ (within $\tau_u$).  Moreover, we have
to retriangulate a suitable sub-pseudo-triangle $\tau_0$ of $\tau_u$
whose boundary, according to Proposition \ref{Thm:LogCostEvent}, has
expected complexity $O(\log n)$ (see also Figure \ref{Fig:EnvelopeEvent}).  To do so, we first determine $\tau_0$,
by locating the edge $q\nu^*(q)$ in the edge list of $\T(\tau_u)$, and then looking for
the lowest (resp., highest) edge above (resp., below) $q\nu^*(q)$ which is generated by a vertex whose priority is smaller than $\prior(q)$.
We then recursively
triangulate $\tau_0$, as described in the proof of Theorem
\ref{Thm:InitStructure}. All this can be done in $O(\log^2n)$ expected
time.  Therefore, we can process any
$\CE$-certificate failure in $O(\log^2n)$ expected time.

\smallskip
\noindent {\bf \emph{$\CV$-certificates}.} We finally consider the case when a visibility event, involving some point $p$ within the funnel of $\tau_v$,
for some node $v$ of
$\PT^+(P)$, causes the failure of the corresponding $\CV$-certificate at some time $t_0$.  Since the failed certificate is associated with an
internal vertex of $\L(\tau_v)$ or $\R(\tau_v)$, all the necessary
updates are local to the funnel of $\tau_v$, and to its triangulation
$\T(\tau_v)$.  We update the $\CV$-certificate of $p$ and insert its
new failure time into $\Q$, in $O(\log n)$ time (the new neighbor of $\nu(p)$ is easily obtained in $O(\log n)$ from the respective edge list). In addition, we may
have to determine and re-triangulate a suitable sub-pseudo-triangle
$\tau_0$ of $\tau_v$, whose boundary has expected complexity $O(\log
n)$ (see Proposition \ref{Thm:LogCostEvent}). An in the case of a failure of a $\CE$-certificate, this can be done in $O(\log^2 n)$ expected time, by searching the edge list of $\T(\tau_v)$.

We thus obtain our main theorem.
\begin{theorem}\label{MainTheorem}
Let $P(t)$ be a collection of $n$ moving points, as above. We can maintain the triangulation $\T(P(t))$ under motion in a kinetic data structure of linear size, which processes an expected number of $O(n^2\beta_{s+2}(n)\log n)$ events, each in $O(\log^2n)$ expected time, where $s$ is the maximum number of times at which any single triple of points of $P(t)$ can become collinear. 
\end{theorem}

\medskip
\noindent{\bf Enforcing locality.} As implied by Theorem \ref{MainTheorem}, the proposed data structure for maintaining $\T(P)$ is compact, efficient, and responsive (where the last two properties hold in expectation). To make it also local (in expectation), it is sufficient to ensure that at any moment of time the expected number of kinetic certificates involving any single point is $O(\log n)$. Clearly, each point is associated with at most two $\CT$-certificates.
Since the expected depth of $\PT^+(P)$ is $O(\log n)$ and each pseudo-triangle of $\PT^+(P)$ defines a single $\CE$-certificate, each point participates in an expected number of $O(\log n)$ $\CE$-certificates.

We next slightly modify the definition of $\CV$-certificates, in order to ensure that at any moment of time the total expected number of $\CV$-certificates involving any point is also $O(\log n)$.
Consider a fixed moment of time $t_0$ and a fixed node $v$ in $\PT^+(P(t_0))$, and choose any vertex $p$ on, say, the left chain $\L(\tau_v)$.
Currently, $p$ participates in a single certificate that it generates (ensuring the validity of $\nu(p)$), and in an arbitrary number of certificates generated by all the vertices $q$ of $\R(\tau_v)$ satisfying $\nu(q)=p$.
We modify our algorithm by keeping (i.e., storing in $\Q$ the failure times of) only the certificates of $p$ that are generated by the leftmost and the rightmost such vertices $q$ in $\R(\tau_v)$. 
If $p$ lies on $\R(\tau_v)$, we act symmetrically. We apply this modification to every node $v$ and every vertex of $\L(\tau_v)\cup \R(\tau_v)$.
This modification does not affect the correctness of the kinetic data structure because, as can be easily checked, among all the $\CV$-certificates involving $p$ and points $q$ with $\nu(q)=p$, the first to fail must be the extreme ones that we keep.

Now, at each node $v$, every vertex of $\L(\tau_v),\R(\tau_v)$ participates in at most three $\CV$-certificates. Since the expected depth of $\PT^+(P)$ is $O(\log n)$, the asserted (expected) locality bound follows. The kinetic maintenance of this restricted set of $\CV$-certificates resembles that of the original set, with the following minor modification. Each time when we process a visibility event caused by the failure of some $\CV$-certificate, generated by a vertex $p$ at some node $v$, we also have to recompute the $\CV$-certificates involving the old and the new points $\nu(p)$. This can be done in $O(\log n)$ time using a binary search over $\L(\tau_v)$ or $\R(\tau_v)$, which does not affect the time bounds in Theorem \ref{MainTheorem}.

\end{document}